# DYNAMIC MODELLING OF HEPATITIS C VIRUS TRANSMISSION AMONG PEOPLE WHO INJECT DRUGS: A METHODOLOGICAL REVIEW

**Running title: Dynamic modelling of HCV transmission among PWID**

Anthony Cousien[1,2], Viet Chi Tran[3], Sylvie Deuffic-Burban[1,2,4], Marie Jauffret-Roustide[5,6], Jean-Stéphane Dhersin[7], Yazdan Yazdanpanah[1,2,8]

[1]IAME, UMR 1137, INSERM, F-75018 Paris, France

[2]IAME, UMR 1137, Univ Paris Diderot, Sorbonne Paris Cité, F-75018 Paris, France

[3]Laboratoire Paul Painlevé UMR CNRS 8524, UFR de Mathématiques, Université des Sciences et Technologies Lille 1, Cité Scientifique, F-59655, Villeneuve d'Ascq Cedex, France

[4]Inserm U995, Université Lille 2 – Lille Nord de France, Lille, France

[5]CERMES3: Centre de Recherche Médecine, Sciences, Santé, Santé Mentale et Société, UMR CNRS 8211 (INSERM U988 Université Paris Descartes, Ecole des Hautes Etudes en Sciences Sociales), Paris, France

[6]Institut de Veille Sanitaire, Saint-Maurice, France

[7]Université Paris 13, Sorbonne Paris Cité, LAGA, CNRS, UMR 7539, F-93430, Villetaneuse, France

[8]Service des Maladies Infectieuses et Tropicales, Hôpital Bichat Claude Bernard, Paris, France

**Corresponding author**
Anthony Cousien, ATIP-Avenir, Inserm U995, Parc Eurasanté, 152 rue du Professeur Yersin, 59120 Loos, France ; anthony.cousien@gmail.com
Tel: 33 (0)3 20 44 59 62; Fax: 33 (0)3 20 96 86 62

# ABSTRACT

Equipment sharing among people who inject drugs (PWID) is a key risk factor in infection by hepatitis C virus (HCV). Both the effectiveness and cost-effectiveness of interventions aimed at reducing HCV transmission in this population (such as opioid substitution therapy, needle exchange programs or improved treatment) are difficult to evaluate using field surveys. Ethical issues and complicated access to the PWID population make it difficult to gather epidemiological data. In this context, mathematical modelling of HCV transmission is a useful alternative for comparing the cost and effectiveness of various interventions. Several models have been developed in the past few years. They are often based on strong hypotheses concerning the population structure. This review presents compartmental and individual-based models in order to underline their strengths and limits in the context of HCV infection among PWID. The final section discusses the main results of the papers.

# INTRODUCTION

In high income countries, people who inject drugs (PWID) are the main population at risk of infection with hepatitis C virus (HCV), with a seroprevalence ranging between 15% and 90% (1). The risk of HCV transmission is high for all drug-equipment sharing that can lead to blood contact: injection equipment (syringes, cotton or cups (2)), straws (3) and crack pipes (4).

Risk reduction measures have been taken to reduce HIV and HCV transmission among PWID. These measures have focused mainly on opioid substitution treatments such as methadone and buprenorphine, and on needle exchange programs (5). However, other measures are possible. Indeed, some European countries have financed and launched supervised injection and needle exchange programs in prisons (6). Moreover, the landscape of therapy for HCV infection, where treatment was suboptimal until recently, is rapidly changing. More efficient and tolerable treatment strategies have become available. Previr-containing regimens – already available – have significantly increased the chances of a sustained virologic response (SVR, i.e. undetectable levels of HCV for an extended period of time) for patients infected with genotype 1 HCV, the most prevalent HCV genotype in western Europe and North America among PWID (5, 7). Using these combinations, for patients treated for the first time, the SVR rate for genotype 1 reaches 70-75% vs. 50% for treatments with pegylated interferon and ribavirin (8-12). Other, more effective pan-genotypic drugs that are used orally and may be prescribed for a shorter duration are at advanced stages of development (13, 14). Given that non-viremic patients (i.e. those with a SVR) cannot transmit the infection (15), we are now considering the use of treatment as a means of preventing transmission of HCV in this population (« treatment as prevention »).

However, the implementation of harm reduction programs (HRP), like needle exchange programs, and the use of these new treatments to avoid transmission imply additional costs.

For optimal use of available resources, it is important to evaluate the effectiveness and cost-effectiveness of these strategies. For this purpose, clinical studies and traditional epidemiological studies such as historical comparisons, cohorts and/or case-control studies encounter problems of feasibility, cost and time, especially in the PWID population, which is difficult to reach. Mathematical modeling is an alternative; indeed, it enables an estimation of the efficiency and cost of multiple strategies of harm reduction, screening and treatment effects upon HCV transmission within a short period of time. The main goal of the present article is to review mathematical models used to simulate transmission of HCV among PWID, and to evaluate their pros and cons.

The first section briefly summarizes the corpus of papers found. The second describes compartmental models and their properties. The third section describes individual-based models and their properties. The final section presents results obtained and recommendations for public health policies.

## SEARCH STRATEGY AND SELECTION CRITERIA

The aim of the review was to identify dynamic mathematical models used for transmission of HCV among PWID in the literature and to evaluate their strengths and limits. Eligible studies had to satisfy two conditions: 1) describe a dynamic mathematical model for transmission of HCV; and 2) study a population of PWID. In accordance with Cochrane collaboration guidelines (16), we conducted our search in the Medline database using the keywords *mathematical*, *model*, *hepatitis C* and *drug user* and variations of these words. Our review takes into account the bibliographies of all identified publications and includes all items found until June 2014.

# RESULTS

Of 214 articles identified, we examined 57 of them in details and keep 32 articles that fulfilled the above criteria (see Figure 1 for details about the reasons for exclusion). We added 5 articles coming from other sources, given a total of 37 original articles (17-53). The modeled populations were mainly from the United Kingdom (17-22, 26, 35, 36, 43, 45-48), Australia (27, 31, 33, 43, 50, 51) and USA (30, 37-39, 49, 52). Two articles involved analyses in developing countries: that of Durier *et al.* (25) studied a population of PWID in Vietnam, while Vickerman *et al.* (24) examined a population of PWID in Pakistan.

The main objectives of these studies were: (i) to present a model of transmission of HCV and provide analytical results concerning the mathematical properties of the model model (28, 35-39); (ii) to evaluate the impact of HRP (on the chronic/antibody prevalence, incidence, number of infections and quality adjusted life years (QALYs) gained) (21, 22, 24-26, 29, 33, 36-41, 50, 51), (iii) to compare epidemics dynamic of HCV and HIV infections (1, 23, 24, 28, 33, 40-42, 44, 45); (iv) to evaluate the impact of treatment of HCV infection on transmission (17-19, 25, 34, 43, 49, 50) and cost (20, 43); and (v) to evaluate the impact of potential HCV vaccination strategy (30, 32, 51). Objectives and main results of these articles are detailed in Table 1.

Mathematical models used in the articles were divided into two categories: compartmental models and individual-based (or agent-based) models (IBMs). The assumptions underlying the two models and their strengths and weaknesses are presented in the following sections In addition, we distinguished two approaches in these articles: stochastic or deterministic. In the stochastic approach, the epidemic is considered as a chain of events (infections, recovery, etc.) occurring at random times among individuals. Thus, this approach takes into account the randomness of durations of the different health stages in the population. At each event, the transition that occurs is determined by probabilities induced by the global transition rates.

When dealing with large populations, averaging of randomness leads to deterministic evolutions that can be described by ordinary differential equations (54). For further explanation about the advantages and weakness of these two approaches, the reader can refer to (55).

## COMPARTMENTAL MODELS

### Description

Compartmental models were the most frequently used class of models for HCV epidemic simulation among PWID with 31 of the 37 articles in our review (see Table 1). They considered transmission of HCV infection at the macroscopic scale, dividing the population into compartments corresponding to different states of the infection process: susceptible, infectious, recovered, etc. (54, 56). Transitions from one state to another were based on rates that could be time dependent.

For instance, Figure 2 presents a model we have recently developed for HCV transmission in PWID (57). The population is distributed in eight compartments associated with HCV infection and care status. Susceptible PWID are separated into two categories: new and experienced injectors, as recently initiated injectors are at higher risk of infection (58). When infected, they progress to acute hepatitis that can lead either to spontaneous recovery (and PWID become susceptible again) or to undiagnosed chronic hepatitis C. Infected patients could be diagnosed, and progress to diagnosed & non-linked to care hepatitis C stages. They are then linked to care and lost to follow-up, or treated. When treated, patients may respond to treatment and return to a susceptible state, or not respond and progress to a non-SVR state. Transitions between different compartments or states are governed by annual transition rates that may be time-dependent. For example, rates of infection depend on the current proportion of infected PWID in the population.

Other published models include more compartments or less compartments according to the objectives and available data. The most parsimonious models include only two compartments: susceptible and infectious (28, 33, 37-39). Some authors included in their model the uncertain immunization (59) of previously infected PWID against re-infection (18, 19, 21, 24, 35, 45). Durier *et al.* differentiated symptomatic and asymptomatic acute infection (25). Esposito *et al.* and Cipriano *et al.* added compartments to model the use of intravenous drugs as an epidemic linked to the HCV epidemic (29). Martin *et al.* took into account the possibility of imprisonment. Additional compartments can allow including substitution therapies (27, 47, 48). Finally, in order to estimate the cost of the infection or the consequences of HCV infection on the population's health, some authors included additional compartments of natural history of chronic hepatitis C (20, 52).

**Strengths and limits**

Compartmental models have the advantage of not needing intensive computation time. In large populations, the trajectory of the epidemics can be described by a system of ordinary differential equations (55). However, they are based on two important hypotheses.

First, the population of a compartment is supposed to be *homogeneous*. The only characteristic of an individual is the compartment in which he/she is located. Thus, parameters that define an individual are parameters that are derived from the overall population; they can be easily interpreted in term of epidemiology and obtained from epidemiological studies. This is not necessarily the case for individual-based models (IBMs; see below). For example, the infection rate represents the force of infection in the population, for which estimations are available for HCV among PWID (58). However, PWID population is highly heterogeneous. There is a variability of individual characteristics and behaviors, which can impact the risk of infection by HCV or HCV-infected patients care. Indeed, the high variations of infection risk among PWID are due to differences in injection frequency, numbers of syringe sharing and number of sharing partners (60). Moreover, response rates to HCV treatment are different according to HCV genotypes. With a pegylated interferon + ribavirin regimen, SVR is around 75 for naïve patients infected with genotypes 2 or 3; and around 50% for naïve patients infected with genotype 1 (8-12). One may hypothesize that this could lead to the selection of hard-to-treat genotypes, and thus to a large proportion of prevalent cases infected with genotype 1. For simplicity, most of the published models do not differentiate genotypes; and consider only an average duration of treatment and SVR rate in the population. Cipriano *et al.* study represents an exception, differentiating genotypes 1-4 of genotypes 2-3 (52). Nevertheless, with the availability of highly effective pan-genotypic anti-HCV regimens not differentiating genotypes will no more be a problem. Some other characteristics are known to impact HCV transmission or treatment success: time since first injection, as the risk of transmission is higher in recent frequent injectors (58); substitution therapy with reduced risk of transmission from those who receive substitution (61); gender, as the spontaneous recovery rate is higher in women (62); HIV-HCV co-infection, associated with lower treatment success (63), imprisonment, as a history of injection in prison increase the risk of infection (64), and even genetic characteristics, with the IL28B polymorphism impacting the treatment response (65). Stratification of the population into different risk groups is a solution for relaxing the hypothesis of homogeneity of PWID (21-25, 27, 29, 36, 40-45). Vickerman *et al.* considered 3 levels of risk corresponding to no syringe sharing, low- and high-frequency syringe sharing (21-23). Corson *et al.* structured their population by experience at injection: recent injectors were more likely to be infected than experienced injectors (36). Zeiler *et al.* aimed at estimating the impact of methadone maintenance programs and structured their population by methadone intake (27). However, stratification of the population is equivalent to introducing different compartments for each group, which can turn the model highly complex. For example, the model in Cipriano *et al.* has 756 compartments (52). Such models imply to estimate a lot of parameters, and data about PWID are often scarce due to the difficulty to reach this population by epidemiological studies.

Secondly, the infection rate is often based upon the hypothesis that the population is *totally mixed*, in the sense that each susceptible individual can potentially be infected by any infectious person: in classical models, the infection rate per susceptible individual $\beta I(t)$ increases with the number of infected persons in the population. This hypothesis seems to be poorly suited for describing infectious contacts with a blood-borne pathogen among PWID. Indeed, it has been shown that PWID share their injection material with a restricted group of injection partners. Wylie *et al.* found that PWID in Canada have few other PWID in their individual network: a median of 3.5 was found for a period of 30 days (60). Brewer *et al.* found a mean number of 18 injecting partners (not necessarily involving syringe/needle sharing) during a 12-month period of presumed HCV infection among HCV-positive PWID in Seattle (66). Sacks-Davis *et al.* reported, in Melbourne, a median number of 3 injection partners/IDU, with a median duration of 3 years for a partnership, and they found that HCV phylogeny was associated with the injection network (67). These results suggest that the number of potential infectious contacts is restricted to a small group for each infected PWID, which slows the transmission of the virus in the population. However, they are based on a snapshot, and give no indication at a lifetime scale. In addition, Hahn *et al.* suggest that because of the high number of injecting partners the turnover of the injecting partners may be sufficiently high to consider that a totally mixed population is a valid hypothesis (30). An alternative approach chosen by several authors to relax this hypothesis in compartmental models is to include *assortative mixing*: they varied the mixing of the different risk groups. Fourteen articles in the present review among the 31 compartmental models took into account assortative mixing in the population according to risk groups (21-24, 40, 42-45, 48), substitution therapies (25, 27), experience as injector (46) or age (47). However, this solution does not enable taking into account small subgroups at the individual or community scale (injecting partners groups), but only groups at a population scale.

## INDIVIDUAL-BASED MODELS

### Description

In 2001, Pollack already underlined that, while a simple model enables obtaining several important results, a homogeneous totally mixed population implies biases in the estimation of the effectiveness of HRP and does not enable to assess effectiveness of targeting interventions (37-39). A possibility to overcome these hypotheses is the use of individual-based models (IBMs). IBMs simulate the patients' trajectories at an individual level, so that we can attach to each of them a specific set of characteristics (age, gender, alcohol consumption, frequency of risk-taking, etc.), on which the different transition rates may depend (68-70).

For HCV epidemic modeling among PWID, only a few authors have used IBMs. Mather *et al.* considered a model taking into account isolated groups of individuals with possible immunization of individuals (51). De Vos *et al.* developed a model taking into account HCV and HIV infections in which mortality and transmission of viruses were based on individual characteristics: age, time since the first injection and time since infection (41). Hahn *et al.* developed an IBM to take into account different behaviors and levels of risk exposure in the population (30). The authors distinguished two practices at risk: risky needle sharing (RNS) and ancillary equipment sharing (AES); the probability of infection varied based on the level of risk exposure (corresponding to frequencies of RNS and AES) and HCV stage of the partner (higher infectivity during acute infection). Hutchinson *et al.* and Rolls *et al.* presented IBMs that took into account the social network of the population (see below) (26, 31, 50).

**Modeling contact network**

The network of contacts can be represented by a graph, *i.e.* a set of vertices representing individuals and interconnected by a set of edges representing potentially infectious contacts between them (see example Figure 3).

Various network models have been described in the literature ((71) and therein). The choice of these models depends on the expected characteristics of the network. Some simple graphs which are easily implemented have unrealistic characteristics and neglect major aspects of the network structure, when focusing on the case of the PWID social networks. They are often based only on individual information like the degree, i.e. the number of individuals in contact with that person in the network. For example, the *configuration model* is constructed from a chosen degree distribution (72, 73). Following this distribution, a degree is attributed to each member of the population, by giving half-edges to each individual. The half-edges is then connected to another at random.

Local information, for example the degree, which can be obtained relatively easily using traditional epidemiological studies, by requesting the number of sharing partners or injection partners for each participant. More complex models such as *intersection graph models* (74, 75), *household graph models* (76, 77) and *stochastic block models* (78) have been proposed to reproduce characteristics of real networks, such as transitivity. The transitivity reflects the preferential association of two persons if they have a common relationship: two individuals are more likely to be friends if they have a common friend. In terms of the graph, this results in a large number of triangles in the network. These models are difficult to calibrate due to the need for information on the global topography of the network. We may have to determine the size of the communities and the probability of connection between individuals of the same group and between individuals of different groups, etc. We cannot obtain such information through traditional studies: independent sampling of the individuals only provides information about the neighborhood of the participants. To catch the global topography of the network, we must use specific methodologies such as *chain-referral sampling* used by Friedman *et al.* to study the network of a population of PWID in New York City (79). In such surveys, new participants are recruited by previous participants among partners. The final result is a subgraph of the population's network.

In the case of HCV in PWID, Hutchinson *et al.* used a configuration model (26). The number of partners for the PWID was generated by a geometric probability distribution with parameter $p \in (0,1)$, i.e. $P(X_i = k) = (1-p)^{k-1}p$. This distribution was suggested by data on the distribution of degree in a population of PWID in Glasgow at the beginning of the 90's. Rolls *et al.* applied an IBM to a real network of 258 PWID in Melbourne, obtained by using *chain-referral sampling* (31). This approach enabled the authors to obtain a very realistic network structure (80). However, it was limited to a small sample due to the difficulty in tracing the contact network of different PWID. They compared results of their model with the empirical network and with a fully connected network (equivalent to the totally mixed hypothesis of compartmental models). They found that time to infection was shorter in a fully connected network, indicating that the structure of the network highly impacts output in HCV transmission. In more recent papers, they calibrated an exponential random graph model (ERGM) on these data (50, 80). ERGMs are statistical models aiming to reproduce some characteristics of the initial network (i.e.; the number of edges, isolates, triangles, in particular). The ERGM used by Rolls *et al.* is based on the research of homophily in the network, which is the preferential attachment between PWID due to their similar characteristics: location, gender, age, frequency of drug use. This is an important part of the social component of the network.

Another research question in PWID is the dynamic of the network. Indeed, in a real population of PWID, the contact pattern may change: the identities of the different sharing partners change over time as relationships between PWID evolve (26). A static network model, with identities of sharing partners fixed over time, may miss a crucial aspect of the social network. It may be a good approximation if contacts change at a slow rate relative to epidemic dynamics. Inversely, if contacts change quickly relatively to the spread of the disease, then the dynamic effects cannot be neglected (81). For HIV epidemics on sexual contact network the dynamic of the sexual partnership network has been proved to impact the epidemic trajectory (82-84). For HCV in PWID, it is an open question. Only Hutchinson *et al.* have used a dynamic network model with a turnover of sharing partners for each individual each year. Two variants of turnover were tested: a turnover totally at random, and a turnover with some stability in injecting partners groups, with few differences between the results in these two settings. These dynamic models are however not be supported by data.

From another point of view, Hahn *et al.* suggest that because of a high dynamic in the contact network, the population can be modelled as static and totally mixed (30). There is no evidence about this affirmation because it is only supported by the high number of reported injecting partners.

Further investigations should be conducted to clarify the necessity of dynamic models. Of note Sacks-Davis *et al.* reported, in Melbourne, a median number of 3 injection partners/IDU, with a median duration of 3 years for a partnership (67), suggesting a relative stability of the network.

## **MAIN RESULTS OF THE REVIEWED PAPERS**

The objectives of articles on HCV transmission among PWID are numerous. However, important results consistently emerge.

**Vaccination**

The potential effect of a vaccination against HCV was evaluated in three articles. Mather *et al.* showed that even immunizing half the population with a vaccine which has an 80% efficacy slows the spread of the infection in an Australian PWID population (51). For Dontwi *et al.*, potential vaccination must be carried out early and should cover a large portion of the population (>80%) so as to reduce the force of infection (32). Hahn *et al.* tested scenarios that specifically target certain individuals in a population of PWID in USA (30). They found that vaccination is most effective when it targets high-risk individuals (with more frequent risk-taking) without taking into account serological status, or when targeting HCV-seronegative persons.

**Harm reduction policies**

The first author to take an interest in estimating the impact of harm reduction by a model was Pollack, who examined a needle exchange program in USA (37-39). His results suggested that, while such a program may have an impact and could potentially eradicate the infection, the cost quickly becomes prohibitive for highly transmissible infections like HCV (>250.000$ per averted infection). However, he suggested that the association of this program with a methadone maintenance therapy strategy would reduce transmission of the virus in the population and, as a consequence, decrease the cost of needle exchange programs. He stated that, for HCV, a combination of different strategies is necessary to impact HCV transmission (39). However, he underlined that the absence of heterogeneous behavior and the totally-mixed hypothesis could impact the results. Vickerman *et al.* also found that widespread sustained coverage of syringe exchange in the population (reduction > 40% of needle/syringe sharing) is necessary in order to obtain a significant reduction in prevalence after 10 years in Rawalpindi, Pakistan (24).

Hutchinson *et al.* estimated the impact of HRP on HCV transmission, by varying the percentage of PWID who had shared per year, the mean number of needle/sharing partners and the percentage of injecting episodes (26). For the period 1988-2000, they estimated that 4,500 infections would have been prevented in Glasgow with HRP. They also found that reducing the mean number of partners to one (*versus* between 2 and 3 partners in the baseline scenario) might prevent 5,300 infections during the same period (with this measure alone). Moreover, wide and sustained decrease of needle/syringe sharing would be necessary to have a similar impact on transmission (5,200 infections prevented): only 11% to 20% of PWID should be able to share a needle/syringe during that period.

Some authors suggested that the target of these interventions could be optimized. Vickerman *et al.* suggested targeting recent injectors not reached by HRP and not already infected in UK to reduce syringe sharing in this particular part of the population (21). They found that a significant reduction in seroprevalence could occur among recent PWID (<4 years) with a reduction in sharing frequency <25%, although among experienced PWID (>8 years), similar results would only occur with a reduction >50%. Similarly, Corson *et al.* suggested that interventions in Scotland are most efficient during the first 5 years of the injecting career (36). Esposito *et al.* in Italy showed a delay of one year between the peak of drug use and the peak of prevalence for HCV, suggesting also that interventions should occur early during the injecting drug career to impact HCV transmission (29).

In the context of limited resources, De Vos *et al.* suggested that PWID at low risk (i.e. less frequent syringe sharing) should be targets for HRP (aiming at reducing syringe sharing rate) so as to maximize their impact (41). They questioned also whether the decrease in HCV incidence in Amsterdam since 1990 was related to HRP. They found that realistic results with their model could only be obtained in the presence of HRP, but demographic changes in the PWID population primarily explained the decrease.

**Impact of harm reduction strategies on HIV and HCV infection among PWID**

The impact of HRP upon HIV transmission is much stronger than upon HCV transmission (33). Coutin *et al.* and Pollack showed that the higher infectivity of HCV compared to HIV implied that greater effort is needed to significantly impact HCV transmission in their model (28, 39). Vickerman *et al.* estimated that reducing the injection risk by 30% would result in a reduction in the incidence/seroprevalence of 50%/28% for HIV and 37%/10% for HCV after 5 years in a population of PWID in UK (23). De Vos *et al.* found similar results (42). Murray *et al.* estimated the number of annual injection partners below which infections by material sharing were less likely to occur than infections by other sources to be 17 partners/year for HIV and 3 partners/year for HCV in an Australian setting (33). They estimated the actual number of partner to be intermediate (≈6), which explains the success obtained against HIV and more questionable results for HCV. Bayoumi *et al.* estimated the effectiveness and cost-effectiveness of supervised injection facilities in Vancouver (53). Their results suggest between $14 and $18 million gained and between 920 and 1175 life-years saved after 10 years. However, the authors underlined that the cost saving is mainly due to HIV infections averted.

**HCV treatment**

The impact of HCV treatment on transmission is considered to be effective despite the risk of reinfection (17, 19, 25, 34). Some authors recommended specific targets and application modalities. Zeiler *et al.* studied the impact of HCV treatment in Australia, taking into account methadone maintenance therapy (27). The main concern of that paper was to determine the optimal distribution of treatment in the population. The results suggested treating active PWID rather than PWID under methadone maintenance therapy, with the hypothesis of equal adherence to treatment in the two groups. The conclusion would be reversed only if adherence by active PWID was <44% of that of PWID under methadone maintenance therapy. Also in Australia, Hellard *et al.* found that even a modest annual rate of treatment (25/1,000 PWID) could have an impact on long-term HCV chronic prevalence (50% decrease after 30 years)(34). Martin *et al.* found similar results for the United Kingdom: for a baseline chronic prevalence of 20%, 40% and 60%, an annual treatment rate of 10/1,000 PWID would achieve a reduction in chronic prevalence of 31%, 13% and 7%, respectively after 10 years (19). Durier *et al.,* in Vietnam (with few HRP) found similar results (25). They estimated a strategy of treatment as prevention, and suggested treating early (during the first year) to avoid a maximum of infections.

Martin *et al.* estimated the cost-effectiveness of HCV treatment in the United Kingdom (20). Their results showed that treating active PWID and ex- or non-PWID was cost-effective, but for a chronic prevalence below 60%; treating active PWID was more cost-effective because of avoided re-infection. This result remained valid even with a SVR rate in active PWID that was 50% lower than that of ex- or non-PWID (which may reflect lower adherence to treatment).

Martin *et al.* estimated the impact of a coming direct-acting antiviral, with treatment of shorter duration and with a higher tolerability (43). They compared efficiency and cost in three different geographic settings: Edinburgh, Melbourne and Vancouver. The study showed that halving the chronic prevalence after 15 years would be extremely expensive, particularly in Melbourne and Vancouver (around $50 million) where the chronic prevalence is high.

Finally, in Australia, Rolls *et al.* targeted infected individual for treatment initiation according to their neighbors on the social network, with strategies including random treatment delivery, priority by node degree, treatment of the primary contacts of infected nodes, treatment of primary and secondary contacts of infected nodes, and treatment of treatment of the primary contact of uninfected nodes (50). They found that strategies including the treatment of primary and secondary contacts of some infected PWID randomly chosen for treatment ("ring" treatment) are the most effective strategies for a similar number of treatment starts.

**CONCLUSION**

To date, several different models have been used to study transmission of HCV among PWID. Most of them were built to answer a specific question taking into account only the characteristics of PWID that pertained to that question, and thus averaging the non-relevant characteristics. Specific points seem to recurrently emerge in these articles: the long-term effects of HRP and HCV treatment on HCV prevalence, the advantage of specifically targeting more risky PWID (recent injectors, active PWID or PWID not on methadone maintenance therapy), and the importance of implementing these measures early (at the beginning of the injecting career for HRP, and at the beginning of chronic infection for treatment). However, more general models are needed to compare a combination of different strategies of risk reduction (needle exchange programs, substitution therapy), screening and treatment (efficacy of new treatments). Mathematical modeling can enable evaluating the cost associated with those different strategies and guiding optimal resource allocation.

Most models are compartmental and rely on strong assumptions. An individual-based approach could be an interesting alternative allowing, for example, to estimate the impact of strategies based on PWID characteristics and social networks. The most advanced works on this topic are that of Rolls *et al.* showing that treatment strategies based on the social network have more impact on transmission than a random treatment distribution (50). But such studies still few in numbers. One of the main difficulties lies in the lack of data for first calibrating and next to be used in the model. Data are difficult to obtain in particular regarding the risk of HCV transmission during an exchange, the frequency of material sharing, and social networks and their dynamics over time. As pointed out by Kretzschmar *et al.,* the construction of such models therefore requires multidisciplinary collaboration that includes clinical (transmission risk, treatment efficacy), epidemiological (current state of infection, screening, treatment), mathematical (modeling) and sociological (social network characteristics among PWID) components (85).

## ACKNOWLEDGMENTS

We would like to express our gratitude to Camille Pelat for her helpful advices during the elaboration of this review; and to David A. Rolls for transmitting us the data (adjacency matrix of the graph) used for the construction of Figure 3. We also would like to thank Jerri Bram for linguistic revisions of the text.

## STATEMENT OF INTERESTS

This study was funded by the French Agence Nationale de Recherche sur le Sida et les Hépatites virales (ANRS, http://www.anrs.fr), grant number 95146. This funding source had no involvement in the writing of the manuscript or the decision to submit it for publication. SDB has received grants from Roche, Janssen-Cilag and Schering-Plough, and received consultancy honoraria from Abbvie, Gilead, Janssen, Merck and GlaxoSmithKline. YY received travel grants, honoraria for presentations at workshops and consultancy honoraria from Abbott, Bristol-Myers Squibb, Gilead, Merck, Roche, Tibotec and ViiV Healthcare. None of the other authors report any association that might pose a conflict of interest.

## ABBREVIATIONS

HCV: Hepatitis C virus

HRP: Harm reduction policies

IBM: Individual Based Model

IDU: injecting drug user

SVR: Sustained virological response

QALY: Quality adjusted life year

# REFERENCES

1       Vickerman P, Hickman M, May M, Kretzschmar M, Wiessing L. Can hepatitis C virus prevalence be used as a measure of injection-related human immunodeficiency virus risk in populations of injecting drug users? An ecological analysis. *Addiction*. 2010; 105(2):311-8.

2       Mathei C, Shkedy Z, Denis B*, et al.* Evidence for a substantial role of sharing of injecting paraphernalia other than syringes/needles to the spread of hepatitis C among injecting drug users. *J Viral Hepat*. 2006; 13(8):560-70.

3       Aaron S, McMahon JM, Milano D*, et al.* Intranasal transmission of hepatitis C virus: virological and clinical evidence. *Clin Infect Dis*. 2008; 47(7):931-4.

4       Jauffret-Roustide M, Le Strat Y, Couturier E*, et al.* A national cross-sectional study among drug-users in France: epidemiology of HCV and highlight on practical and statistical aspects of the design. *BMC Infect Dis*. 2009; 9:113.

5       Trepo C, Pradat P. Hepatitis C virus infection in Western Europe. *J Hepatol*. 1999; 31 Suppl 1:80-3.

6       Jauffret-Roustide M, Pedrono G, Beltzer N. Supervised consumption rooms: The French Paradox. *Int J Drug Policy*. 2013; 14(6):628-30.

7       Payan C, Roudot-Thoraval F, Marcellin P*, et al.* Changing of hepatitis C virus genotype patterns in France at the beginning of the third millenium: The GEMHEP GenoCII Study. *J Viral Hepat*. 2005; 12(4):405-13.

8       Hezode C, Forestier N, Dusheiko G*, et al.* Telaprevir and peginterferon with or without ribavirin for chronic HCV infection. *N Engl J Med*. 2009; 360(18):1839-50.

9       Kwo PY, Lawitz EJ, McCone J*, et al.* Efficacy of boceprevir, an NS3 protease inhibitor, in combination with peginterferon alfa-2b and ribavirin in treatment-naive patients with genotype 1 hepatitis C infection (SPRINT-1): an open-label, randomised, multicentre phase 2 trial. *Lancet*. 2010; 376(9742):705-16.

10      McHutchison JG, Everson GT, Gordon SC*, et al.* Telaprevir with peginterferon and ribavirin for chronic HCV genotype 1 infection. *N Engl J Med*. 2009; 360(18):1827-38.

11      Poordad F, McCone J, Bacon B*, et al.* Boceprevir for untreated chronic HCV genotype 1 infection. *N Engl J Med*. 2011; 364(13):1195-206.

12      Sherman K, Flamm S, Afdhal N*, et al.* Response-guided telaprevir combination treatment for hepatitis C virus infection. *N Engl J Med*. 2011; 365(11):1014-24.

13      Bourliere M, Khaloun A, Wartelle-Bladou C*, et al.* Chronic hepatitis C: Treatments of the future. *Clin Res Hepatol Gastroenterol*. 2011; 35 Suppl 2:S84-95.

14      Lawitz E, Poordad FF, Pang PS*, et al.* Sofosbuvir and ledipasvir fixed-dose combination with and without ribavirin in treatment-naive and previously treated patients with genotype 1 hepatitis C virus infection (LONESTAR): an open-label, randomised, phase 2 trial. *Lancet*. 2014; 383(9916):515-23.

15      Yazdanpanah Y, De Carli G, Migueres B*, et al.* Risk factors for hepatitis C virus transmission to health care workers after occupational exposure: a European case-control study. *Clin Infect Dis*. 2005; 41(10):1423-30.

16      Higgins JPT, Green S (editors). Cochrane Handbook for Systematic Reviews of Interventions Version 5.1.0 [updated March 2011]. The Cochrane Collaboration, 2011. Available from www.cochrane-handbook.org.

17      Martin NK, Pitcher AB, Vickerman P, Vassall A, Hickman M. Optimal control of hepatitis C antiviral treatment programme delivery for prevention amongst a population of injecting drug users. *PLoS One*. 2011; 6(8):e22309.

18      Martin NK, Vickerman P, Foster GR, Hutchinson SJ, Goldberg DJ, Hickman M. Can antiviral therapy for hepatitis C reduce the prevalence of HCV among injecting drug user populations? A modeling analysis of its prevention utility. *J Hepatol*. 2011; 54(6):1137-44.

19	Martin NK, Vickerman P, Hickman M. Mathematical modelling of hepatitis C treatment for injecting drug users. *J Theor Biol*. 2011; 274(1):58-66.
20	Martin NK, Vickerman P, Miners A*, et al.* Cost-effectiveness of hepatitis C virus antiviral treatment for injection drug user populations. *Hepatology*. 2012; 55(1):49-57.
21	Vickerman P, Hickman M, Judd A. Modelling the impact on Hepatitis C transmission of reducing syringe sharing: London case study. *Int J Epidemiol*. 2007; 36(2):396-405.
22	Vickerman P, Martin N, Turner K, Hickman M. Can needle and syringe programmes and opiate substitution therapy achieve substantial reductions in HCV prevalence? Model projections for different epidemic settings. *Addiction*. 2012; 107(11):1984-95.
23	Vickerman P, Martin NK, Hickman M. Understanding the trends in HIV and hepatitis C prevalence amongst injecting drug users in different settings-Implications for intervention impact. *Drug Alcohol Depend*. 2011; 123(1-3):122-31.
24	Vickerman P, Platt L, Hawkes S. Modelling the transmission of HIV and HCV among injecting drug users in Rawalpindi, a low HCV prevalence setting in Pakistan. *Sex Transm Infect*. 2009; 85 Suppl 2:ii23-30.
25	Durier N, Nguyen C, White LJ. Treatment of hepatitis C as prevention: a modeling case study in Vietnam. *PLoS One*. 2012; 7(4):e34548.
26	Hutchinson SJ, M.B.; S, A.; T, S.J.; G. Modelling the spread of hepatitis C virus infection among injecting drug users in Glasgow : Implications for prevention. *Int J Drug Policy*. 2006; 17:211-21.
27	Zeiler I, Langlands T, Murray JM, Ritter A. Optimal targeting of Hepatitis C virus treatment among injecting drug users to those not enrolled in methadone maintenance programs. *Drug Alcohol Depend*. 2010; 110(3):228-33.
28	Coutin L, Descreusefond L, Dhersin JS. A Markov model for the spread of viruses in an open population. *J Appl Prob*. 2010; 47:976-96.
29	Esposito N, Rossi C. A nested-epidemic model for the spread of hepatitis C among injecting drug users. *Math Biosci*. 2004; 188:29-45.
30	Hahn JA, Wylie D, Dill J*, et al.* Potential impact of vaccination on the hepatitis C virus epidemic in injection drug users. *Epidemics*. 2009; 1(1):47-57.
31	Rolls DA, Daraganova G, Sacks-Davis R*, et al.* Modelling hepatitis C transmission over a social network of injecting drug users. *J Theor Biol*. 2011; 297C:73-87.
32	Dontwi IK, Frempong NK, Bentil DE, Adetunde I, Owusu-Ansah E. Mathematical modeling of Hepatitis C Virus transmission among injecting drug users and the impact of vaccination. *Am J Sci Ind Res*. 2010; 1(1):41-6.
33	Murray JM, Law MG, Gao Z, Kaldor JM. The impact of behavioural changes on the prevalence of human immunodeficiency virus and hepatitis C among injecting drug users. *Int J Epidemiol*. 2003; 32(5):708-14.
34	Hellard ME, Jenkinson R, Higgs P*, et al.* Modelling antiviral treatment to prevent hepatitis C infection among people who inject drugs in Victoria, Australia. *Med J Aust*. 2012; 196(10):638-41.
35	Corson S, Greenhalgh D, Hutchinson S. Mathematically modelling the spread of hepatitis C in injecting drug users. *Math Med Biol*. 2012; 29(3):205-30.
36	Corson S, Greenhalgh D, Hutchinson SJ. A time since onset of injection model for hepatitis C spread amongst injecting drug users. *J Math Biol*. 2012; 66(4-5):935-78.
37	Pollack HA. Ignoring 'downstream infection' in the evaluation of harm reduction interventions for injection drug users. *Eur J Epidemiol*. 2001; 17(4):391-5.
38	Pollack HA. Cost-effectiveness of harm reduction in preventing hepatitis C among injection drug users. *Med Decis Making*. 2001; 21(5):357-67.
39	Pollack HA. Can we protect drug users from hepatitis C? *J Policy Anal Manage*. 2001; 20(2):358-64.
40	De Vos AS, Kretzschmar ME. The efficiency of targeted intervention in limiting the spread of HIV and Hepatitis C Virus among injecting drug users. *J Theor Biol*. 2013; 333:126-34.

41	de Vos AS, van der Helm JJ, Matser A, Prins M, Kretzschmar ME. Decline in incidence of HIV and hepatitis C virus infection among injecting drug users in Amsterdam; evidence for harm reduction? *Addiction*. 2013; 108(6):1070-81.

42	de Vos AS, van der Helm JJ, Prins M, Kretzschmar ME. Determinants of persistent spread of HIV in HCV-infected populations of injecting drug users. *Epidemics*. 2012; 4(2):57-67.

43	Martin NK, Vickerman P, Grebely J*, et al.* HCV treatment for prevention among people who inject drugs: Modeling treatment scale-up in the age of direct-acting antivirals. *Hepatology*. 2013; 58(5):1598-1609.

44	Castro Sanchez AY, Aerts M, Shkedy Z*, et al.* A mathematical model for HIV and hepatitis C co-infection and its assessment from a statistical perspective. *Epidemics*. 2013; 5(1):56-66.

45	Vickerman P, Martin NK, Roy A*, et al.* Is the HCV-HIV co-infection prevalence amongst injecting drug users a marker for the level of sexual and injection related HIV transmission? *Drug Alcohol Depend*. 2013; 132(1-2):172-81.

46	Corson S, Greenhalgh D, Taylor A, Palmateer N, Goldberg D, Hutchinson S. Modelling the prevalence of HCV amongst people who inject drugs: An investigation into the risks associated with injecting paraphernalia sharing. *Drug Alcohol Depend*. 2013; 133(1):172-9.

47	Martin NK, Hickman M, Miners A, Hutchinson SJ, Taylor A, Vickerman P. Cost-effectiveness of HCV case-finding for people who inject drugs via dried blood spot testing in specialist addiction services and prisons. *BMJ Open*. 2013; 3(8):e003153.

48	Martin NK, Hickman M, Hutchinson SJ, Goldberg DJ, Vickerman P. Combination interventions to prevent HCV transmission among people who inject drugs: modeling the impact of antiviral treatment, needle and syringe programs, and opiate substitution therapy. *Clin Infect Dis*. 2013; 57 Suppl 2:S39-45.

49	Elbasha EH. Model for hepatitis C virus transmissions. *Math Biosci Eng*. 2013; 10(4):1045-65.

50	Rolls DA, Sacks-Davis R, Jenkinson R*, et al.* Hepatitis C transmission and treatment in contact network of people who inject drugs. *PLoS One*. 2013; 8(11):e78286.

51	Mather D, Crofts N. A computer model of the spread of hepatitis C virus among injecting drug users. *Eur J Epidemiol*. 1999; 15(1):5-10.

52	Cipriano LE, Zaric GS, Holodniy M, Bendavid E, Owens DK, Brandeau ML. Cost effectiveness of screening strategies for early identification of HIV and HCV infection in injection drug users. *PLoS One*. 2012; 7(9):e45176.

53	Bayoumi AM, Zaric GS. The cost-effectiveness of Vancouver's supervised injection facility. *CMAJ*. 2008; 179(11):1143-51.

54	Andersson H, Britton T. *Stochastic epidemic models and their statistical analysis*. New York: Springer 2000.

55	Ball F. Dynamic population epidemic models. *Math Biosci*. 1991; 107(2):299-324.

56	Kermack W, McKendrick A. A Contribution to the Mathematical Theory of Epidemics. *Proceedings of the Royal Society of London Series A*. 1927; 115:700-21.

57	Cousien A, Tran VC, Jauffret-Roustide M, Deuffic-Burban S, Dhersin J-S, Yazdanpanah Y. O89 Impact of new DAA-containing regimens on HCV transmission among injecting drug users (IDUs): a model-based analysis (ANRS 12376). *Journal of Hepatology*. 2014; 60(1):S36-S7.

58	Sutton AJ, Gay NJ, Edmunds WJ, Hope VD, Gill ON, Hickman M. Modelling the force of infection for hepatitis B and hepatitis C in injecting drug users in England and Wales. *BMC Infect Dis*. 2006; 6:93.

59	Mehta SH, Cox A, Hoover DR*, et al.* Protection against persistence of hepatitis C. *Lancet*. 2002; 359(9316):1478-83.

60	Wylie JL, Shah L, Jolly AM. Demographic, risk behaviour and personal network variables associated with prevalent hepatitis C, hepatitis B, and HIV infection in injection drug users in Winnipeg, Canada. *BMC Public Health*. 2006; 6:229.

61	Backmund M, Reimer J, Meyer K, Gerlach JT, Zachoval R. Hepatitis C virus infection and injection drug users: prevention, risk factors, and treatment. *Clin Infect Dis*. 2005; 40 Suppl 5:S330-5.
62	van den Berg CH, Grady BP, Schinkel J*, et al.* Female sex and IL28B, a synergism for spontaneous viral clearance in hepatitis C virus (HCV) seroconverters from a community-based cohort. *PLoS One*. 2011; 6(11):e27555.
63	Graham CS, Baden LR, Yu E*, et al.* Influence of human immunodeficiency virus infection on the course of hepatitis C virus infection: a meta-analysis. *Clin Infect Dis*. 2001; 33(4):562-9.
64	Sutton AJ, McDonald SA, Palmateer N, Taylor A, Hutchinson SJ. Estimating the variability in the risk of infection for hepatitis C in the Glasgow injecting drug user population. *Epidemiol Infect*. 2012; 140(12):2190-8.
65	Ge D, Fellay J, Thompson AJ*, et al.* Genetic variation in IL28B predicts hepatitis C treatment-induced viral clearance. *Nature*. 2009; 461(7262):399-401.
66	Brewer DD, Hagan H, Sullivan DG*, et al.* Social structural and behavioral underpinnings of hyperendemic hepatitis C virus transmission in drug injectors. *J Infect Dis*. 2006; 194(6):764-72.
67	Sacks-Davis R, Daraganova G, Aitken C*, et al.* Hepatitis C virus phylogenetic clustering is associated with the social-injecting network in a cohort of people who inject drugs. *PLoS One*. 2012; 7(10):e47335.
68	Bonabeau E. Agent-based modeling: methods and techniques for simulating human systems. *Proc Natl Acad Sci U S A*. 2002; 99 Suppl 3:7280-7.
69	Epstein JM, Axtell R, 2050 Project. *Growing artificial societies : social science from the bottom up*. Washington, D.C.: Brookings Institution Press 1996.
70	Gillespie DT. A general method for numerically simulating the stochastic time evolution of coupled chemical reactions. *Journal of Computational Physics*. 1976; 22:403-34.
71	Newman MEJ. The structure and function of complex networks. *SIAM*. 2003; 45(2):167-256.
72	Molloy M, Reed B. A critical point for random graphs with a given degree sequence. *Random Struct Algor*. 1995; 6(2-3):161-80.
73	Bollobás B. Degree sequences of random graphs. *Discrete Mathematics*. 1981; 33(1):1-19.
74	Singer K. Random intersection graphs. Vol. PhD thesis: Johns Hopkins University, 1995.
75	Karonski M, Scheinerman E, Singer-Cohen K. On random intersection graphs: the subgraphs problem. *Combinatorics, Probability & Computing*. 1999; 8:131-59.
76	Bartoszynski R. On a certain model of an epidemic. *Zastos Mat*. 1972/73; 13:139–51.
77	Becker NG, Dietz K. The effect of household distribution on transmission and control of highly infectious diseases. *Math Biosci*. 1995; 127:207–19.
78	Holland PW, Laskey KB, Leinhardt S. Stochastic blockmodels: Some first steps. *Social Networks*. 1983; 5:109-37.
79	Friedman SR, Neaigus A, Jose B*, et al.* Sociometric risk networks and risk for HIV infection. *Am J Public Health*. 1997; 87(8):1289-96.
80	Rolls DA, Wang P, Jenkinson R*, et al.* Modelling a disease-relevant contact network of people who inject drugs. *Social Networks*. 2013; 35(4):699-710.
81	Volz E, Meyers LA. Susceptible-infected-recovered epidemics in dynamic contact networks. *Proc Biol Sci*. 2007; 274(1628):2925-33.
82	Kretzschmar M, Dietz K. The effect of pair formation and variable infectivity on the spread of an infection without recovery. *Mathematical biosciences*. 1998; 148(1):83-113.
83	Kim J-H, Koopman JS. HIV transmissions by stage in dynamic sexual partnerships. *Journal of theoretical biology*. 2012; 298:147-53.
84	Shafer LA, Nsubuga RN, Chapman R, O'Brien K, Mayanja BN, White RG. The dual impact of antiretroviral therapy and sexual behaviour changes on HIV epidemiologic trends in Uganda: a modelling study. *Sexually transmitted infections*. 2014;90(5):423-9.
85	Kretzschmar M, Wiessing L. New challenges for mathematical and statistical modeling of HIV and hepatitis C virus in injecting drug users. *AIDS*. 2008; 22(13):1527-37.

**Table 1:** Summary of the review

| Reference | Country (setting) | Objectives | Model/approach | Main results |
|---|---|---|---|---|
| Mather *et al.* (51) | Australia | To obtain predicted outcomes of different scenarios of immunization by an hypothetical vaccine and to identify sensitive parameters of the model | Individual-based model / stochastic approach | The rate of spread of HCV trough an PWID population is not really sensitive to the initial prevalence, but it is sensitive to the proportion of groups members with whom the individual has contact and the probability of infection per contact with an infective (encouraging HRP)<br><br>A hypothetical vaccine with 80% efficacy would have a measurable impact on the spread of the infection, even with moderate coverage rate (around 50%) |
| Pollack (39) | USA | To assess the impact of a syringe exchange program on HCV so as to understand why it is effective for HIV but not for HCV in an PWID population | Compartmental model/ deterministic differential equations - Analytic results (steady state analysis) | Modest interventions (reduce incidence by a third) are only effective for hard-to-transmit infections (like HIV), while for HCV, with a high prevalence, only a massive program will have an impact. Moreover, for a high-prevalence setting, the impact will occur over the long term |
| Pollack (38) | USA | To assess the effectiveness and cost-effectiveness of a syringe exchange program | Compartmental model/ deterministic differential equations - Analytic results (steady-state analysis) | The syringe exchange program is effective and cost-effective when R0 is low (2.9 for HIV), but becomes less effective when R0 is high (6.9 for HCV) and cost become prohibitive (>250.000$ per averted infection), which is more realistic (in short-term analysis). This is necessary to combine syringe exchange program with intervention aimed at decreasing the R0, such as the methadone maintenance program. |

| Reference | Country (setting) | Objectives | Model/approach | Main results |
|---|---|---|---|---|
| Pollack (37) | USA | To assess biases that represent short-term analysis compared to a long-term analysis (i.e. taking into account changes in prevalence in steady-state analysis) in the evaluation of harm reduction intervention among PWID | Compartmental model/ deterministic differential equations - Analytic results (steady-state analysis) | The short-term incidence underestimates the effectiveness of a program of long-term syringe exchange if the steady-state prevalence, in the absence of intervention, is below 50%. Conversely, if it is over 50%, the short-term incidence overestimates effectiveness. |
| Murray *et al.* (33) | Australia | To assess the impact of different levels of needle sharing on the prevalence of HIV and HCV among PWID | Compartmental model/ deterministic differential equations | The needle exchange program has less impact on HCV than on HIV: the number of annual injection partners below which infections by material sharing were less likely to occur than infections by other sources to be 17 partners/year for HIV and 3 partners/year for HCV |
| Esposito *et al.* (29) | Italy | To present a mathematical model taking into account the "epidemic" of drug use | Compartmental model/ deterministic differential equations | HRP must be initiated early to be efficient<br><br>Incidence is the best indicator of impact (delays are shorter than for prevalence) |
| Hutchinson *et al.* (26) | United-Kingdom | To model the hepatitis epidemic among PWID in Glasgow and implications for prevention | Individual-based model on a simulated contact network of PWID / stochastic approach | The estimated seroprevalence in the 90s was lower than currently observed. This result can be corrected by increasing viremia during the acute phase, which could confirm the higher rate of viremia during that period.<br><br>Current public health messages (not sharing needles) are inadequate for HCV (but effective for HIV). The authors propose encouraging PWID to share with only a small group of trusted persons (tested negative) and predict that by reducing the mean number of partners to one (vs. 2-3) 5,300 infections could have been avoided between 1988 and 2000 |

| Reference | Country (setting) | Objectives | Model/approach | Main results |
|---|---|---|---|---|
| Vickerman *et al.* (21) | United-Kingdom | To assess the impact of decreased needle sharing on HCV transmission | Compartmental model/ deterministic differential equations | Current public health interventions aiming at reducing needle/syringe sharing (like needle/syringe exchange) should target new or recently initiated injectors to have an optimal impact. There is therefore a need to work on new injectors which is currently not the case.<br><br>Moreover, we have to target all PWID, not just those with a high frequency of syringe sharing<br><br>Frequency sharing must be significantly reduced (to 1-2 times a month) to achieve a seroprevalence ≤ 10% |
| Bayoumi *et al.* (53) | Canada | To estimate the cost-effectiveness of Vancouver's supervised injection facility | Compartmental model/ deterministic differential equations | If a reduction of needle sharing is the only effet of the supervised injection facility: incremental net savings of almost $14 million and 920 life-years saved after 10 years.<br><br>When adding an increase use of safe injection pratices and increased referral to methadone maintenance treatment: incremental net savings of almost $18 million and 1175 life-years saved after 10 years.<br><br>The cost saving is mainly due to HIV infections averted. |
| Hahn *et al.* (30) | USA | To assess the impact of vaccinations (at different levels of effectiveness) on the HCV epidemic among PWID | Individual-based model / stochastic approach | Several scenarios of vaccination were tested:<br>- randomly vaccinate the population<br>- vaccinate individuals with more risky behavior<br>- vaccinate individuals seronegative for HCV (sero-targeting)<br>The third scenario is the most effective, followed by the second, suggesting that random vaccination is not a good strategy. |

| Reference | Country (setting) | Objectives | Model/approach | Main results |
|---|---|---|---|---|
| Vickerman *et al.* (24) | Pakistan | To explore different hypotheses to explain the low prevalence of HCV among PWID in Rawalpindi, Pakistan and to estimate the impact of interventions | Compartmental model/ deterministic differential equations | Most syringe sharing involves low risk, because it concern only a small group of the users' acquaintances - Existence of a small group that carries out high-risk sharing with strangers, in which the prevalence is high<br><br>Predicted increase in HIV prevalence in 5-10 years<br><br>A reduction of syringe sharing > 40% could reduce the number of HCV/HIV infections of around 45% after 10 years if all PWID are reached |
| Zeiler *et al.* (27) | Australia | To study the impact of HCV treatment allocation according to methadone taking | Compartmental model/ deterministic differential equations | Advantage of treating active users rather than users on methadone because of re-infection and high turnover of PWID on methadone. |
| Dontwi *et al.* (32) | Unspecified | To assess the impact of a possible vaccine against HCV among PWID | Compartmental model/ deterministic differential equations | Potential vaccination carried out early and covering a large part of the population (>80%) would significantly reduce the force of infection (nearly 90%) and, ultimately, the extent of the epidemic |
| Coutin *et al.* (28) | Unspecified | To present a mathematical model for the spread of a virus in an open population such as HCV and HIV, and evaluate sensitivity to parameters | Compartmental model/ deterministic differential equations and stochastic approach (with analytical study of the convergence of the stochastic model to the deterministic model) | A quarantine (i.e. isolate the population to prevent transmission) ensures a long-term decline in HCV prevalence but is impractical<br><br>Increasing the number of PWID leads to a decrease in HCV prevalence (also impractical)<br><br>The differing force of infection explain the different results between HIV and HCV |

| Reference | Country (setting) | Objectives | Model/approach | Main results |
|---|---|---|---|---|
| Martin *et al.*(18) | United-Kindgom | To study the impact of treatment on chronic prevalence of hepatitis C among PWID | Compartmental model/ deterministic differential equations | Treatment has a significant impact on transmission of HCV in the population despite the risk of reinfection |
| Martin *et al.* (19) | United-Kindgom | To assess the level of treatment required to eradicate or control the epidemic of hepatitis C among PWID | Compartmental model/ deterministic differential equations | Treatment has a significant impact on transmission of HCV in the population despite risk of re-infection, even for low treatment rates (<6% of chronically infected annually for chronic prevalence <40%; 10-20% for a chronic prevalence of 60%) |
| Martin *et al.* (17) | United-Kindgom | To optimize the number of treatments taking into account economic constraints | Compartmental model/ deterministic differential equations | An increase in the annual budget allocated to treatment would be cost-effective and would more rapidly reduce chronic prevalence. |
| Vickerman *et al.* (23) | Unspecified | To understand the trends in HIV and hepatitis C seroprevalence among PWID in different settings | Compartmental model/ deterministic differential equations | Existence of a threshold for the seroprevalence of HCV, below which HIV prevalence is negligible This threshold depends on the environment (practices, etc.) |
| | | | | The existence of different levels of risk groups and the size of these groups could explain the range of observed values for the prevalence of HIV / HCV in different settings |
| | | | | Strategies for long-term intervention needed to reduce the seroprevalence of HCV |
| Rolls *et al.* (31) | Australia | To propose a model of transmission of HCV among PWID | Individual-based model on an empirical contact network of PWID / stochastic approach | Re-infection rates (20.6/100 PY) are higher than rates of primary infection (14.4/100 PY) Comparison with a fully connected graph (equivalent to the assumption of compartmental models) does not achieve this |

| Reference | Country (setting) | Objectives | Model/approach | Main results |
|---|---|---|---|---|
| | | | | Event transmission rate estimated: 1% |
| Cipriano *et al.* (52) | USA | To estimate the cost, effectiveness and cost-effectiveness of HCV and HIV screening (antibody and/or viral RNA testing) for PWID in OST | Compartmental model/ deterministic differential equations | Depending on screening frequency; adding HIV/HCV viral RNA testing to antibody testing adverts between 14.8 and 30.3 HIV infections and between 3.1 and 7.7 HCV infections in a population of 26,100 screened PWID entering in OST.<br><br>Strategies including HCV testing have incremental cost-effectiveness ratio > \$100,000/QALY gained, unless awareness of HCV infection status results in a decrease >5% of needle sharing. |
| Castro Sanchez *et al.* (44) | Italy | To choose a model, identifying parameters to which the model is sensitive, fitting the model | Compartmental model/ deterministic differential equations | Selection of a model (from two evaluated models) for the force of infection; and identification of the number of risk groups in the population (two: low and high risk).<br><br>The most sensitive parameters are those linked to syringe sharing and transmission rates in chronic stages of HIV and HCV infection |
| Corson *et al.* (35) | United-Kingdom | To present a mathematical model for the spread of HCV in PWID<br><br>To determine the level of needle or syringe sharing, needle cleaning or needle exchange necessary for an eventual elimination of HCV | Compartmental model/ deterministic differential equations | The model predicts $R_0 < 1$ and thus an eventual elimination of HCV infection for one of the following situations:<br>- syringe sharing rate $\leq 54·67$/year<br>- needle cleaning $\geq 0·74$<br>- needle turnover $\geq 562·37$/year |

| Reference | Country (setting) | Objectives | Model/approach | Main results |
|---|---|---|---|---|
| Corson et al. (36) | United-Kingdom | To present a mathematical model for the spread of HCV in PWID<br><br>To study the basic reproductive number ($R_0$)<br><br>To study the impact of needle exchange | Compartmental model/ deterministic differential equations | Demonstration that with the authors model for $R_0 \leq 1$: tends toward elimination of HCV infection; meanwhile for $R_0 > 1$: unique endemic equilibrium distribution. In Glasgow, $R_0$ is estimated to be 3.613.<br><br>The interventions are more efficient if they target recent PWID (<5 years of injection) |
| Martin et al. (20) | United-Kingdom | To estimate the cost-effectiveness of HCV therapy among PWID | Compartmental model/ deterministic differential equations | Treating active injchronicectors and non-injectors is cost-effective, but if HCV chronic prevalence is below 60%, it is more cost-effective to treat active injectors |
| Durier et al. (25) | Vietnam | To estimate the preventive effect of HCV therapy, methadone maintenance therapy and needle/syringes exchanges programs in a developing country context | Compartmental model/ deterministic differential equations | Even a low level of treatment (25%; 4 years into infection) has a significant impact on chronic prevalence (reduction >21% after 11 years)<br><br>Adding needle/syringes exchanges programs and substitution treatment in greater numbers provides an additional gain<br><br>Advantage of implementing measures to diagnose patients at earlier stages of the disease ("Treatment as Prevention") |
| Vickerman et al. (22) | United-Kingdom | To investigate the impact of scaling-up opiate substitution therapy (OST) and high coverage needle and syringe programs on HCV chronic prevalence | Compartmental model/ deterministic differential equations | Scaling-up opiate substitution therapy and needle sharing programs can reduce hepatitis C chronic prevalence among PWID, but reductions may be modest and require long-term sustained intervention coverage: in UK, to reduce the chronic prevalence from 40% to less than 30% over 10 years needs a coverage $\geq 80\%$) |

| Reference | Country (setting) | Objectives | Model/approach | Main results |
|---|---|---|---|---|
| Hellard et al. (34) | Australia | To estimate the effect of HCV treatment on HCV chronic prevalence among PWID | Compartmental model/ deterministic differential equations | Modest rates of current HCV treatment among PWID in Victoria, Australia (25 per 1000 PWID) could halve HCV chronic prevalence in 30 years |
| De Vos et al. (42) | Netherlands | To understand the dynamics of HCV and HIV infection among PWID | Compartmental model/ deterministic differential equations | The link between HIV and HCV prevalence is linked to distribution of risk, and assortativity of groups of risk. There is a threshold for HCV prevalence below which HIV does not spread |
| De Vos et al. (41) | Netherlands | To understand the effect of HRP on the decline in the incidence of HCV and HIV among PWID since 1990 | Individual-based model/ stochastic approach | It is difficult to reproduce realistic behavior of epidemics without HRP; however, most of the decline can be explained by demographic changes |
| De Vos et al. (40) | Netherlands | To estimate the effectiveness of targeted intervention on HIV and HCV among PWID | Compartmental model/ deterministic differential equations | HRP are most effective toward HIV if used in a high-risk group, but they must be used in a low-risk group to impact HCV. High-risk individuals are already infected by HCV (the prevalence is high) so it is too late to prevent their infection, while this is not the case for HIV (low prevalence due to lower force of infection) |
| Vickerman et al. (45) | United-Kingdom | To understand the link between HIV/HCV co-infections and the HIV sexual transmission rate in a population of PWID | Compartmental model/ deterministic differential equations | To reproduce the realistic seroprevalence of HCV among HIV-infected PWID, it is necessary to include sexual transmission. Moreover, the level of co-infection seems to be a marker of HIV sexual transmission among PWID |

| Reference | Country (setting) | Objectives | Model/approach | Main results |
|---|---|---|---|---|
| Martin *et al.* (43) | United-Kingdom Australia Canada | To estimate the effectiveness of future direct-acting antivirals on HCV chronic prevalence among PWID in 3 different settings (Edinburgh, Melbourne and Vancouver) | Compartmental model/ deterministic differential equations | The impact will be limited by current treatment coverage. To halve the chronic prevalence of HCV within 15 years, the cost would be high, especially in Melbourne and Vancouver (~$50 million), where the chronic prevalence is highest |
| Corson *et al.* (46) | United-Kingdom | To understand the role of injecting paraphernalia (filters, cookers and water) | Compartmental model/ deterministic differential equations | The transmission probability is estimated to be at least 8 times lower through paraphernalia- than through needle- or syringe-sharing. Paraphernalia sharing is estimated to significantly contribute to HCV infections (62% of HCV infections in Scotland with current estimated needle/syringe sharing rates and paraphernalia sharing rates) |
| Martin *et al.* (47) | United-Kingdom | To estimate the cost-effectiveness of HCV case-findings for PWID via dried blood spot (DBS) testing in addiction services and prisons | Compartmental model/ deterministic differential equations | For a £20,000 per QALY gained willingness-to-pay threshold, DBS testing is cost-effective in addiction services, but not in prison. If we increase continuity of care (proportion of initiated treatments/referrals that are continued when entering/exiting prison) to 40%, DBS testing become effective in prison. |
| Martin *et al.* (48) | United-Kingdom | To estimate the impact of combining opiate substitution therapy, high-coverage needle and syringe exchange programs and HCV treatment on chronic prevalence and incidence | Compartmental model/ deterministic differential equations | HCV treatment is necessary to achieve a large reduction (>45%) in HCV chronic prevalence over 10 years. Opiate substitution therapy, high-coverage needle and syringe exchange programs and new direct-acting antivirals should reduce the number of necessary treatments. |

| Reference | Country (setting) | Objectives | Model/approach | Main results |
|---|---|---|---|---|
| Elbasha (49) | USA | To assess the impact of treatment on transmission of HCV in an PWID population | Compartmental model/ deterministic differential equations | The incidence can increase or decrease with treatment according to the re-infection rate, but the prevalence is always lower with treatment than without |
| Rolls *et al.* (50) | Australia | To investigate the effect of the number of contacts on time to primary infection and the role of spontaneously clearing nodes on incidence rates; and the effect of treatment strategies based on networks properties on incidence rates of primary infections and reinfections | Individual-based model on a simulated contact network of PWID / stochastic approach | The number of contacts and injecting frequency play a key role in reducing the time before primary infection<br><br>The spontaneous clearance has a local effect (i.e. around the concerning individual) on infection risk and the total number of spontaneous recovery has a global effect on the incidence of both primary and re-infection rates<br><br>Network-based treatment strategies that chose PWID and treat their contact are most effective and allow to reduce the number of treatment needed to achieve a desired effect |

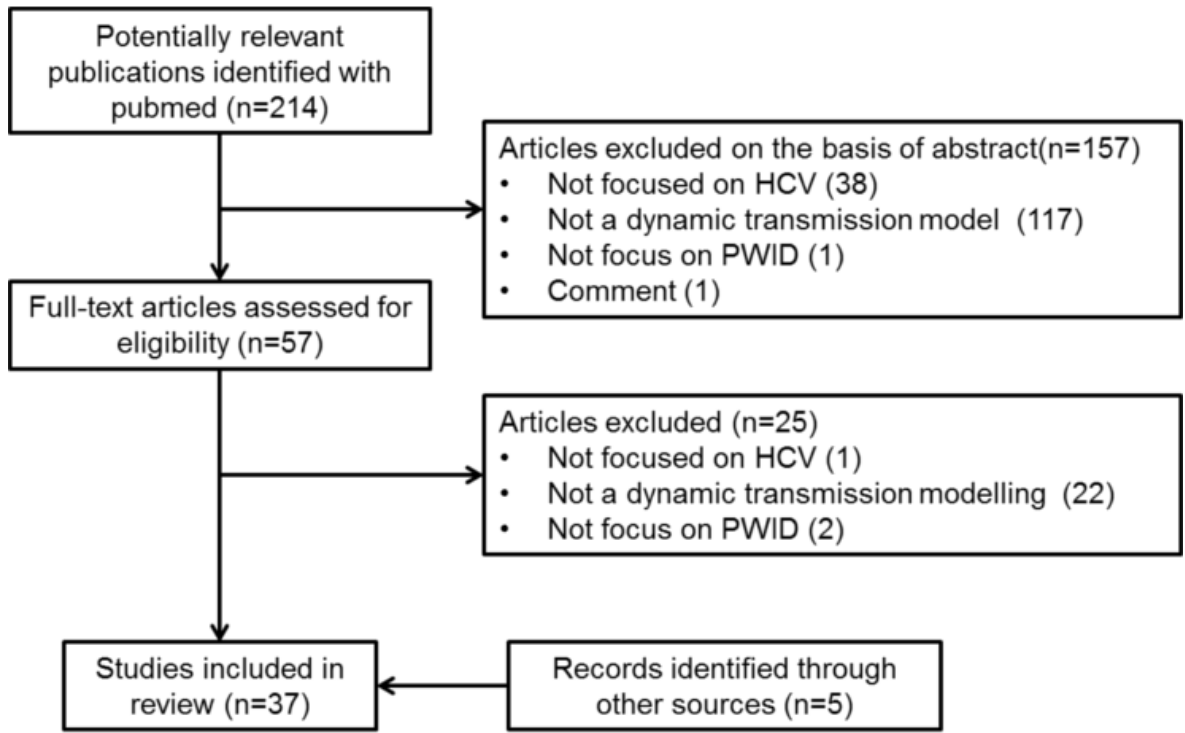

**Figure 1**: Identification of the relevant articles.

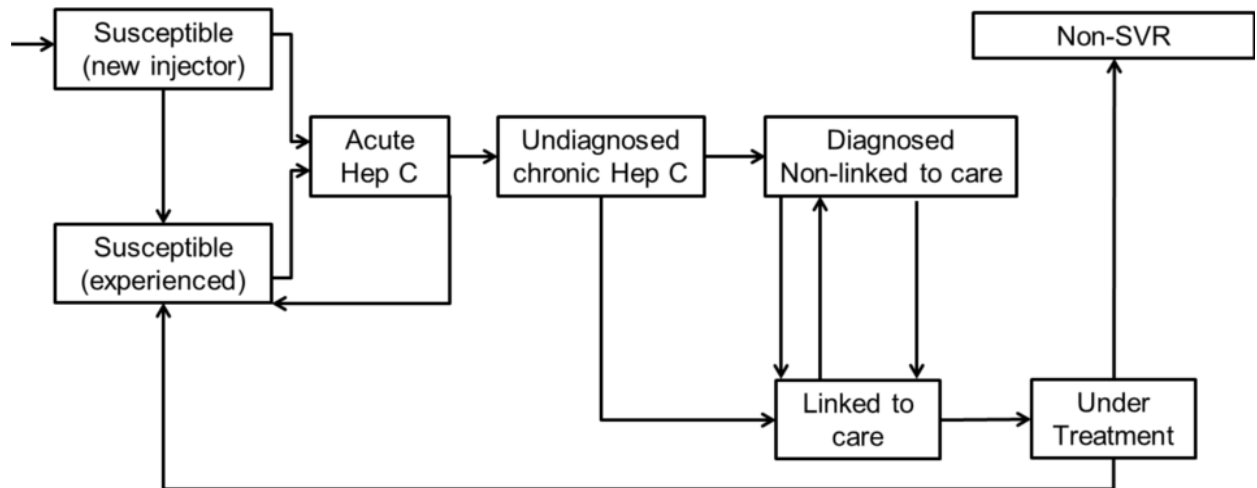

**Figure 2:** Schematic representation of a compartmental model for the transmission of HCV in PWID (57). The model takes into account infectious status, acute hepatitis C, testing, linkage to care, loss to follow-up and treatment. We distinguished recently initiated PWID and experienced PWID because of their different risk of infection (58). Reinfections are possible, but no retreatment is allowed in the model. Transitions occur between the compartments at transition rates that can depend on time: the infection rate depends on the current number of infected PWID in the population.

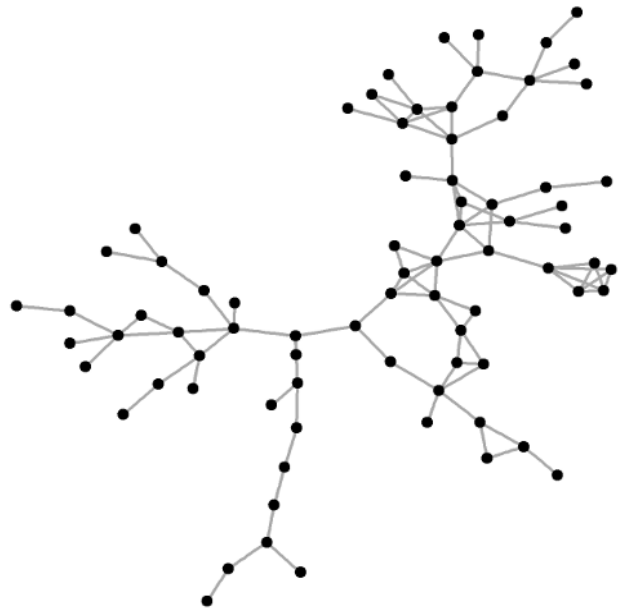

**Figure 3**: Main component of an empirical network of PWID in Melbourne from Rolls *et al.* (31). Each node represents a PWID, and a tie is drawn between the injecting partners in the previous 3 months. Data courtesy of DA Rolls.